\documentclass[12pt]{article}
\usepackage{epsf}
\def\beq{\begin{equation}}
\def\eeq{\end{equation}}
\def\bea{\begin{eqnarray}}
\def\eea{\end{eqnarray}}

\def\nnb{\nonumber}

\def\nnb{\nonumber}

\def\ba{\begin{array}}
\def\ea{\end{array}}
\def\bea{\begin{eqnarray}}
\def\eea{\end{eqnarray}}

\title{ {\bf
$B_{s}\rightarrow  \tau^+ \tau^-$  decay in the general two Higgs doublet
model}}
\author{\vspace{1cm}\\
         {\bf E. O. Iltan} 
         \thanks{E-mail address:
        eiltan@heraklit.physics.metu.edu.tr}
 \\
         Physics Department, Middle East Technical University \\
         Ankara, Turkey\\        \vspace{5mm}\\
        {\bf G. Turan}
        \thanks{E-mail address:
        gsevgur@rorqual.metu.edu.tr}
 \\
        Physics Department, Middle East Technical University \\
        Ankara, Turkey\\}

\date{}

\begin{document}
\setlength{\baselineskip}{24pt}
\maketitle
\setlength{\baselineskip}{7mm}
\begin{abstract}
We study  the exclusive decay 
$B_s\rightarrow \tau^+ \tau^-$ in the general two Higgs doublet model.
We analyse the dependencies of the branching ratio
 on the  model parameters, including the leading order QCD 
corrections. We found that there is an enhancement in the branching ratio,
especially for $r_{tb}=\frac{\bar{\xi}_{N,tt}^U}{\bar{\xi}_{N,bb}^D}>1$ case. 
Further, the neutral Higgs effects 
are detectable for large values of the parameter $\bar{\xi}_{N,\tau\tau}^D$.
\end{abstract} 
\thispagestyle{empty}
\newpage
\setcounter{page}{1}
\section{Introduction}
The study of rare B-decays is one of the most important research areas in
particle physics and there is an experimental effort for studying them at
various centers such as SLAC (BaBar), KEK (BELLE), B-Factories, DESY 
(HERA-B). In the Standard model (SM) they are induced by flavor changing 
neutral currents (FCNC) at loop level and therefore they are sensitive to 
the fundamental parameters, like Cabbibo-Kobayashi-Maskawa (CKM) matrix
elements, leptonic decay constants, etc. These decays also provide a
sensitive test to the new physics beyond the SM, such as two Higgs Doublet 
model (2HDM), Minimal Supersymmetric extension of the SM (MSSM) 
\cite{Hewett}, etc.   

Among the rare B decays, $B_s\rightarrow l^+ l^-$ process, induced by the
inclusive  $b\rightarrow s l^+ l^-$ decay, is attractive since the only
non-perturbative quantity in the theoretical calculation is the decay
constant of $B_s$ which is reliably known. From the experimental point of
view, the measurement of the hadronic decay is easier compared to its
inclusive channel. 

The measurement of upper limit of $B_s\rightarrow \mu^+ \mu^-$ \cite{abe}
\begin{eqnarray}
BR(B_s\rightarrow \mu^+ \mu^-)\le 2.6\, 10^{-6}\,\,,
\end{eqnarray}
stimulated the study of $B_s\rightarrow l^+ l^-$ decays. In the literature, 
this process (\cite{R1}-\cite{R7}) and its inclusive one
$B_s\rightarrow X_s l^+ l^-$ (\cite{R8}-\cite{Chao}) have been investigated 
extensively in the SM, 2HDM and supersymmetric model (SUSY). When $l=e,\mu$, 
the neutral Higgs boson (NHB) effects are safely neglected in the 2HDM 
because they enter in the expressions with the factor 
$m_{e(\mu)}/m_W,\,m_{H^{\pm}}$. However, for $l=\tau$, this factor is not 
negligible and NHB effects can give important contribution. 
In \cite{Chao}, $B\rightarrow X_s\tau^+\tau^-$ process was studied 
in the 2HDM and it was shown that NHB effects are sizable  for large 
values of $tan\beta$. Therefore the main observation of these calculations 
is the enhancement of the branching ratio ($BR$) of these decays for large 
$tan\beta$ values in the 2HDM and minimal supersymmetric model (MSSM), 
especially for $l=\tau$ lepton case. In a recent work \cite{gurer2}, the 
inclusive $b\rightarrow s\tau^+\tau^-$ process has been studied in the 
general 2HDM with real Yukawa couplings and it was found that the $BR$ has 
been enhanced for large values of the parameters $\bar{\xi}^D_{N,\tau\tau}$ 
and $\bar{\xi}^D_{N,bb}$. 
In this work, we study the $B_s\rightarrow \tau^+\tau^-$ decay in the general 
2HDM, so-called model III . Our calculations are based on the results of the 
work \cite{gurer2} for the inclusive $b\rightarrow s \tau^+ \tau^-$ 
decay. Here we include NHB effects and make the full calculation using the 
on-shell renormalization prescription. The investigation of the dependencies 
of the $BR$ on the model parameters, 
namely $\bar{\xi}_{N,bb}^D$ and $\bar{\xi}_{N,\tau\tau}^D$, shows that 
a large enhancement in the $BR$ is possible. 

The paper is organized as follows:
In Section 2, we present the leading order (LO) QCD corrected effective
Hamiltonian  and the corresponding matrix element for the exclusive 
$B_{s}\rightarrow \tau^{+} \tau^{-}$ decay in the framework of the model 
III. Section 
3 is devoted to the analysis of the dependencies of the $BR$ 
on the the Yukawa couplings $\bar{\xi}_{N,bb}^D$, $\bar{\xi}_{N,\tau\tau}^D$ 
and to the discussion of our 
results. In Appendices, we present the operators appearing in the effective 
Hamiltonian and their Wilson coefficients.
\section{The $B_{s}\rightarrow \tau^{+} \tau^{-}$ decay in the framework of 
the model III} 
The general Yukawa interaction in the model III is 
\begin{eqnarray}
{\cal{L}}_{Y}&=&\eta^{U}_{ij} \bar{Q}_{i L} \tilde{\phi_{1}} U_{j R}+
\eta^{D}_{ij} \bar{Q}_{i L} \phi_{1} D_{j R}+
\xi^{U}_{ij} \bar{Q}_{i L} \tilde{\phi_{2}} U_{j R}+
\xi^{D}_{ij} \bar{Q}_{i L} \phi_{2} D_{j R}
 + h.c. \,\,\, ,
\label{lagrangian}
\end{eqnarray}
where $L$ and $R$ denote chiral projections $L(R)=1/2(1\mp \gamma_5)$ and
$\phi_{i}$ for $i=1,2$, are two scalar doublets. Here $\eta^{U,D}_{ij}$, 
$\xi^{U,D}_{ij}$ are the Yukawa matrices and, in general, they have complex 
entries. The choice of scalar Higgs doublets  
\begin{eqnarray}
\phi_{1}=\frac{1}{\sqrt{2}}\left[\left(\begin{array}{c c} 
0\\v+H^{0}\end{array}\right)\; + \left(\begin{array}{c c} 
\sqrt{2} \chi^{+}\\ i \chi^{0}\end{array}\right) \right]\, , 
\nonumber \\ \\
\phi_{2}=\frac{1}{\sqrt{2}}\left(\begin{array}{c c} 
\sqrt{2} H^{+}\\ H^1+i H^2 \end{array}\right) \,\, ,\nonumber
\label{choice}
\end{eqnarray}
with the vacuum expectation values,  
\begin{eqnarray}
<\phi_{1}>=\frac{1}{\sqrt{2}}\left(\begin{array}{c c} 
0\\v\end{array}\right) \,  \, ; 
<\phi_{2}>=0 \,\, , 
\label{cb}
\end{eqnarray}
and the gauge and $CP$ invariant Higgs potential which 
spontaneously breaks  $SU(2)\times U(1)$ down to $U(1)$:
\begin{eqnarray}
V(\phi_1, \phi_2 )&=&c_1 (\phi_1^+ \phi_1-v^2/2)^2+
c_2 (\phi_2^+ \phi_2)^2 \nonumber \\ &+& +
c_3 [(\phi_1^+ \phi_1-v^2/2)+ \phi_2^+ \phi_2]^2
+ c_4 [(\phi_1^+ \phi_1) (\phi_2^+ \phi_2)-(\phi_1^+ \phi_2)(\phi_2^+ \phi_1)]
\nonumber \\ &+& 
c_5 [Re(\phi_1^+ \phi_2)]^2 +
c_{6} [Im(\phi_1^+ \phi_2)]^2 
+c_{7} \, ,
\label{potential}
\end{eqnarray}
where constants $c_i, \, i=1,...,7$, permits us to carry the SM particles in 
the first doublet and the information about the new physics by the second
one. The Yukawa interaction 
\begin{eqnarray}
{\cal{L}}_{Y,FC}=
\xi^{U}_{ij} \bar{Q}_{i L} \tilde{\phi_{2}} U_{j R}+
\xi^{D}_{ij} \bar{Q}_{i L} \phi_{2} D_{j R} + h.c. \,\, .
\label{lagrangianFC}
\end{eqnarray}
describes the Flavor Changing (FC) one beyond the SM. Here, the couplings  
$\xi^{U,D}$ for the charged FC interactions are 
\begin{eqnarray}
\xi^{U}_{ch}&=& \xi^U_{N} \,\, V_{CKM} \nonumber \,\, ,\\
\xi^{D}_{ch}&=& V_{CKM} \,\, \xi^D_{N}  \,\, ,
\label{ksi1} 
\end{eqnarray}
and
\begin{eqnarray}
\xi^{U,D}_{N}&=&(V_L^{U,D})^{-1} \xi^{U,D}\, V_R^{U,D}\,\, , 
\label{ksineut}
\end{eqnarray}
where the index "$N$" in $\xi^{U,D}_{N}$ denotes the word "neutral". 
Notice that $H_1$ and $H_2$ are the mass eigenstates $h^0$ and $A^0$ 
respectively, since no mixing occurs between two CP-even neutral bosons 
$H^0$ and $h^0$ in the tree level, for our choice.

The exclusive $B_s\rightarrow \tau^+\tau^-$ process is induced by the 
inclusive $b\rightarrow s \tau^+ \tau^-$ decay. Therefore we start with  
the effective Hamiltionian of $b\rightarrow s \tau^+ \tau^-$ decay
\bea
{\cal{H}}_{eff} & = & \frac{\alpha G_F}{ \sqrt{2}\, \pi} V_{tb} V_{ts}^*
 \Bigg{\{} C_9^{eff} (\bar s \gamma_\mu P_L b) \, \bar \tau \gamma_\mu \tau
+
C_{10} ( \bar s \gamma_\mu P_L b) \, \bar \tau \gamma_\mu \gamma_5 \tau
\nnb \\
& -& 2 C_7 \frac{m_b}{p^2} (\bar s i \sigma_{\mu \nu} p_\nu P_R b) \bar \tau
\gamma_\mu \tau
 + C_{Q_{1}}(\bar s  P_R b) \bar \tau  \tau +C_{Q_{2}}
(\bar s  P_R b) \bar \tau \gamma_5 \tau \Bigg{\}}~,
\label{hamilton}
\eea
with $p=p_1+p_2$, the sum of the momenta of $\tau^{+}$ and $\tau^{-}$.
Note that, $C_{7},C_{9}$ and $C_{10},$ are the Wilson coefficients 
normalized at the scale $\mu $ and given  in Appendix B. The additional 
Wilson coefficients $C_{Q_1}$ and $C_{Q_2}$ are due to the NHB exchange 
diagrams  (see Appendix B). In calculating the ${\cal{H}}_{eff}$, one 
first integrates out the heavy degrees of freedom, namely $t$ quark, 
$W^{\pm}, \,H^{\pm}, \,H_0, \,H_{1}$ and $H_{2}$ bosons in the present 
case and then obtain the effective theory. Here $H^{\pm}$ denote charged, 
and $H^0$, $H_{1}$ ,$H_{2}$ denote neutral Higgs bosons. Note that 
$H_{1}$ and $H_{2}$ are the same as the mass eigenstates $h^{0}$ and 
$A^{0}$ in the model III respectively. At this stage the QCD corrections 
are added through matching the full  theory with the effective low energy 
one at the high scale $\mu=m_{W}$ and evaluating the Wilson coefficients 
from $m_{W}$ down to the lower scale $\mu\sim O(m_{b})$. 

In the model III, the neutral Higgs particles bring additional contributions 
(see eq.(\ref{NHB})) since the mass of  $\tau$ lepton or related Yukawa 
coupling $\bar{\xi}^{D}_{N,\tau\tau}$ enter into the expressions (see 
\cite{gurer2}). Finally the neutral Higgs boson (NHB) contributions are 
calculated using the on-shell renormalization scheme to overcome the 
logarithmic divergences. Using the renormalization condition 
\begin{eqnarray}
\Gamma_{neutr}^{Ren}(p^2)=\Gamma_{neutr}^{0}(p^2)+\Gamma_{neutr}^{C}=0 ,
\label{rencond}
\end{eqnarray}
the counter term $\Gamma_{neutr}^{C}$ and then the renormalized vertex 
function $\Gamma_{neutr}^{Ren}(p^2)$ is obtained. Here the phrase $neutr$ 
denotes the neutral Higgs bosons $H^0$, $h^0$ and  $A^0$ and  $p$ is the 
momentum transfer.

For the exclusive decay $B_{s}\rightarrow \tau^{+} \tau^{-}$, 
${\cal{H}}_{eff}$ is to be taken between vacuum and 
$|B^{0}_{s}>$ state as $<0|{\cal{H}}_{eff}|B^{0}_{s}>$ and 
this matrix element can be expressed in terms of the $B^{0}_{s}$ decay 
constant $f_{B_{s}}$ using 
\bea
<0|\bar s \gamma_\mu \gamma_5 b|B^{0}_{s}> & = & -i f_{B_{s}}p_{\mu}
\label{ff1} \nonumber \,\, , \\
<0|\bar s \gamma_5 b|B^{0}_{s}> & = & i
f_{B_{s}}\frac{m^2_{B_{s}}}{m_b+m_s} \label{ff2} \nonumber \,\, , \\
<0|\bar s \sigma_{\mu \nu} P_R  b|B^{0}_{s}> & = & 0 . 
\label{ff3} 
\eea
Since $p=p_1+p_2$, the $C^{eff}_{9}$ term in eq.(\ref{hamilton}) gives zero
on contraction with the lepton bilinear, $C_7$ gives zero by
eq.(\ref{ff3}) and the  $C_{10}$ term gets a factor of $2m_{\tau}$ while
the remaining $C_{Q_{1}}$ and $C_{Q_{2}}$ terms get $m_{B_{s}}$, when taking
$m_{B_{s}} \approx m_b+m_s$. Thus the effective Hamiltonian eq. 
(\ref{hamilton}) results in the following decay amplitude for 
$B_{s}\rightarrow \tau^{+} \tau^{-}$
\bea
A & = & \frac{G_F \alpha}{2 \sqrt{2} \pi} \, m_{B_{s}} \, f_{B_{s}} V_{tb}
V^*_{ts} \, \Bigg{[}C_{Q_{1}} \, \bar{\tau} \tau +(C_{Q_{2}}-2
\frac{m_{\tau}}{m_{B_{s}}} C_{10}) \bar{\tau} \gamma_5 \tau \Bigg{]}.
\eea
To calculate the branching ratio we find the square of this amplitude,
then sum over the spins of the $\tau$ leptons and finally integrate over the
phase space. This straightforward calculation gives for the branching ratio
of $B_{s}\rightarrow \tau^{+} \tau^{-}$ 
\bea
BR  & = & \frac{G^2_F \alpha^2}{64 \pi^3} \, m^3_{B_{s}} \tau_{B_{s}}
f^2_{B_{s}} |V_{tb}V^*_{ts}|^2 \sqrt{1-4 \frac{m^2_{\tau}}{m^2_{B_{s}}}}
\Bigg{[}(1-4 \frac{m^2_{\tau}}{m^2_{B_{s}}}) |C_{Q_{1}}|^2+ |C_{Q_{2}}-2
\frac{m_{\tau}}{m_{B_{s}}} C_{10}|^2\Bigg{]}. \,\, 
\eea 
%

\section{Discussion}
In the multi-Higgs doublet models, there are many free parameters, such as 
masses of charged and neutral Higgs bosons and the Yukawa
couplings. In the present work  we study our process in the general 2HDM, 
so called model III. The Yukawa couplings, which are entries of Yukawa 
matrices can be restricted using the experimental measurements. In our 
calculations, we neglect all Yukawa couplings except $\bar{\xi}^{U}_{N,tt}$, 
$\bar{\xi}^{D}_{N,bb}$, $\bar{\xi}^D_{N,\tau\tau}$ by respecting the CLEO 
measurement \cite{cleo2}, 
\begin{eqnarray}
BR (B\rightarrow X_s\gamma)= (3.15\pm 0.35\pm 0.32)\, 10^{-4} \,\, .
\label{br2}
\end{eqnarray}

This section is devoted to  $\frac{\bar{\xi}^{D}_{N,bb}}{m_b}$ 
and $\bar{\xi}^{D}_{N,\tau\tau}$ dependencies of $BR$  for the 
exclusive decay $B_s\rightarrow  \tau^+  \tau^-$, restricting $|C_7^{eff}|$ 
in the region $0.257 \leq |C_7^{eff}| \leq 0.439$ due to the CLEO measurement, 
eq.(\ref{br2}) (see \cite{alil1} for details). In our  numerical calculations, 
we take the charged Higgs mass $m_{H^{\pm}}=400\, GeV$ and the scale 
$\mu=m_b$. Further, we use the redefinition
\bea
\xi^{U,D}=\sqrt{\frac{4 G_F}{\sqrt{2}}} \bar{\xi}^{U,D}\,\, , 
\nnb
\eea
 and the input values given in Table (\ref{input}). 
\begin{table}[h]
        \begin{center}
        \begin{tabular}{|l|l|}
        \hline
        \multicolumn{1}{|c|}{Parameter} & 
                \multicolumn{1}{|c|}{Value}     \\
        \hline \hline
        $m_{\tau}$                   & $1.78$ (GeV) \\
        $m_c$                   & $1.4$ (GeV) \\
        $m_b$                   & $4.8$ (GeV) \\
        $m_{H^{0}}$                   & $150$ (GeV) \\
        $m_{h^{0}}$                   & $80$ (GeV) \\
        $m_{A^{0}}$                   & $80$ (GeV) \\ 
        $\alpha_{em}^{-1}$      & 129           \\
        $\lambda_t$            & 0.04 \\
        $m_{t}$             & $175$ (GeV) \\
        $m_{W}$             & $80.26$ (GeV) \\
        $m_{Z}$             & $91.19$ (GeV) \\
        $\Lambda_{QCD}$             & $0.225$ (GeV) \\
        $\alpha_{s}(m_Z)$             & $0.117$  \\
        $sin\theta_W$             & $0.2325$  \\
        \hline
        \end{tabular}
        \end{center}
\caption{The values of the input parameters used in the numerical
          calculations.}
\label{input}
\end{table}

\begin{figure}[htb]
\vskip -3.0truein
\centering
\epsfxsize=6.8in
\leavevmode\epsffile{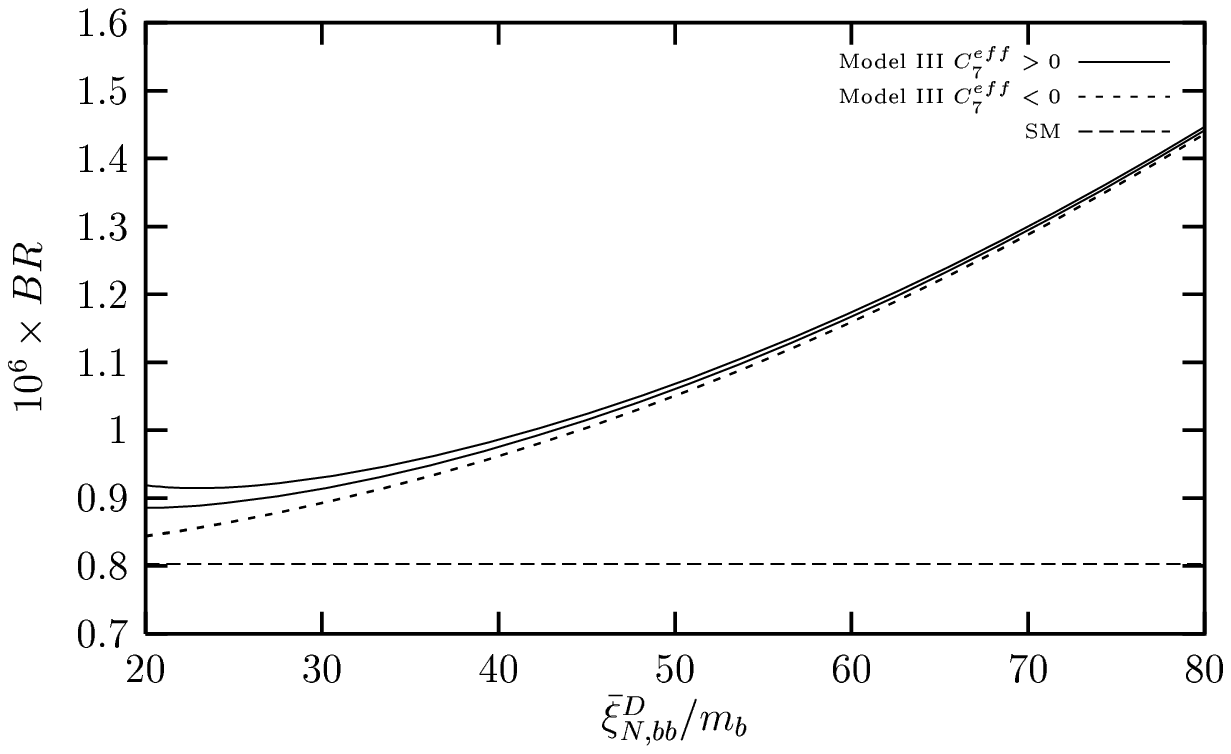}
\vskip -3.0truein
\caption[]{$BR$ as a function of  $\bar{\xi}_{N,bb}^{D}/m_b$ for 
$\bar{\xi}_{N,\tau\tau}^{D}=200\, GeV$ in case of the ratio $|r_{tb}|<1$.}
\label{brksibbRk1}
\end{figure}

\begin{figure}[htb]
\vskip -3.0truein
\centering
\epsfxsize=6.8in
\leavevmode\epsffile{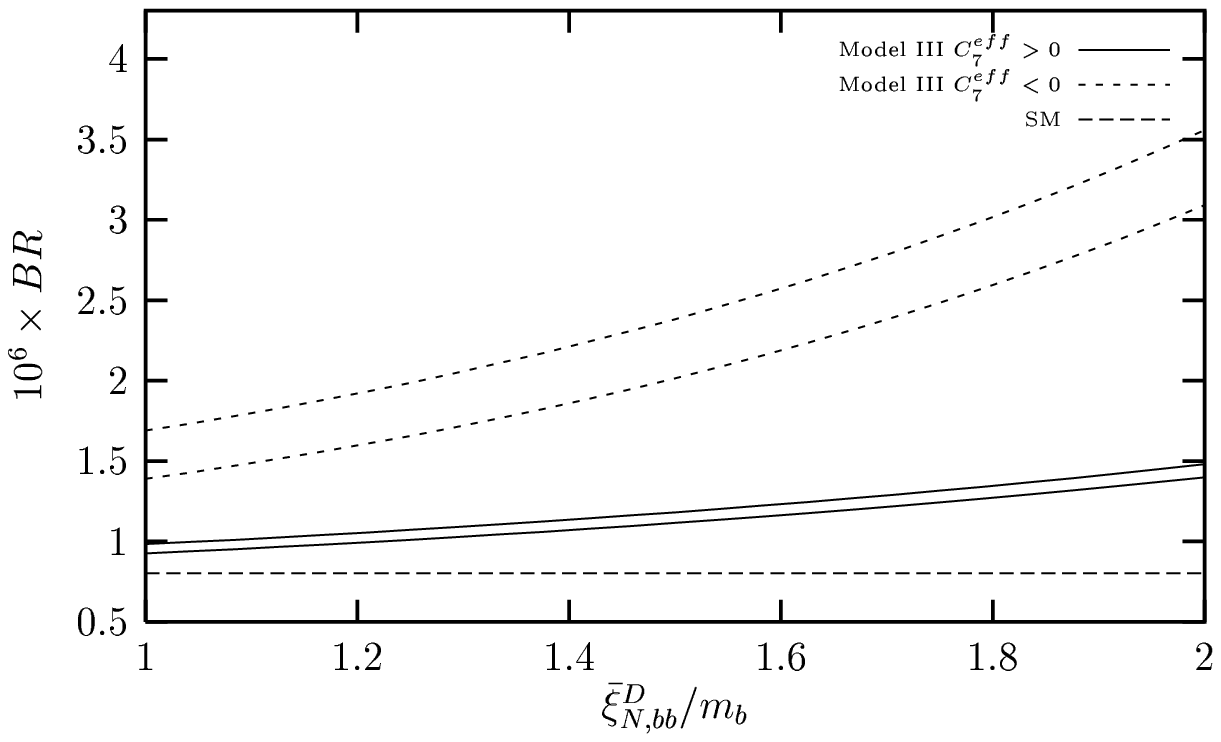}
\vskip -3.0truein
\caption[]{$BR$ as a function of  $\bar{\xi}_{N,bb}^{D}/m_b$ for 
$\bar{\xi}_{N,\tau\tau}^{D}=20\, GeV$ in case of the ratio $r_{tb}>1$.}
\label{brksibbRb1}
\end{figure}

Fig. \ref{brksibbRk1} shows $\frac{\bar{\xi}_{N,bb}^{D}}{m_b}$ dependence of 
the $BR$ of the decay under consideration for $\bar{\xi}_{N,\tau\tau}^{D}
=200 \, GeV$ and the ratio $|r_{tb}|=
|\frac{\bar{\xi}_{N,bb}^{D}}{\bar{\xi}_{N,tt}^{U}}|<1$. The $BR$ is
restricted to the region bounded by solid lines for $C^{eff}_{7}>0$ 
or to the small dashed line for $C^{eff}_{7}<0$. This quantity is 
sensitive to $\frac{\bar{\xi}_{N,bb}^{D}}{m_b}$ and it increases by an 
amount $\%\, 60$ in the interval $20\leq \frac{\bar{\xi}_{N,bb}^{D}}{m_b}
\leq 80$. Besides, the enhancement 
compared to the SM case is predicted as beeing $\% 80$. 
The $\frac{\bar{\xi}_{N,bb}^{D}}{m_b}$
dependence of the $BR$ for $r_{tb}>1$ is presented  in
Fig. \ref{brksibbRb1}. For this case, the $BR$ increases considerably  
even for small values of $\bar{\xi}_{N,\tau\tau}^{D}$, which is taken  $20 \, 
GeV$, in this calculation. The $BR$ enhances with increasing 
$\bar{\xi}_{N,\tau\tau}^{D}$, especially for $C^{eff}_{7}<0$
case. This figure shows that the $BR$ is strongly sensitive to  
the parameter $\bar{\xi}_{N,bb}^{D}$ for $r_{tb}>1$ and it may get the
values four (two) times larger compared to the ones in the SM 
for $C^{eff}_{7}<0$ ($C^{eff}_{7}>0$) even at $\bar{\xi}_{N,bb}^{D}=2\, m_b$.

\begin{figure}[htb]
\vskip -3.0truein
\centering
\epsfxsize=6.8in
\leavevmode\epsffile{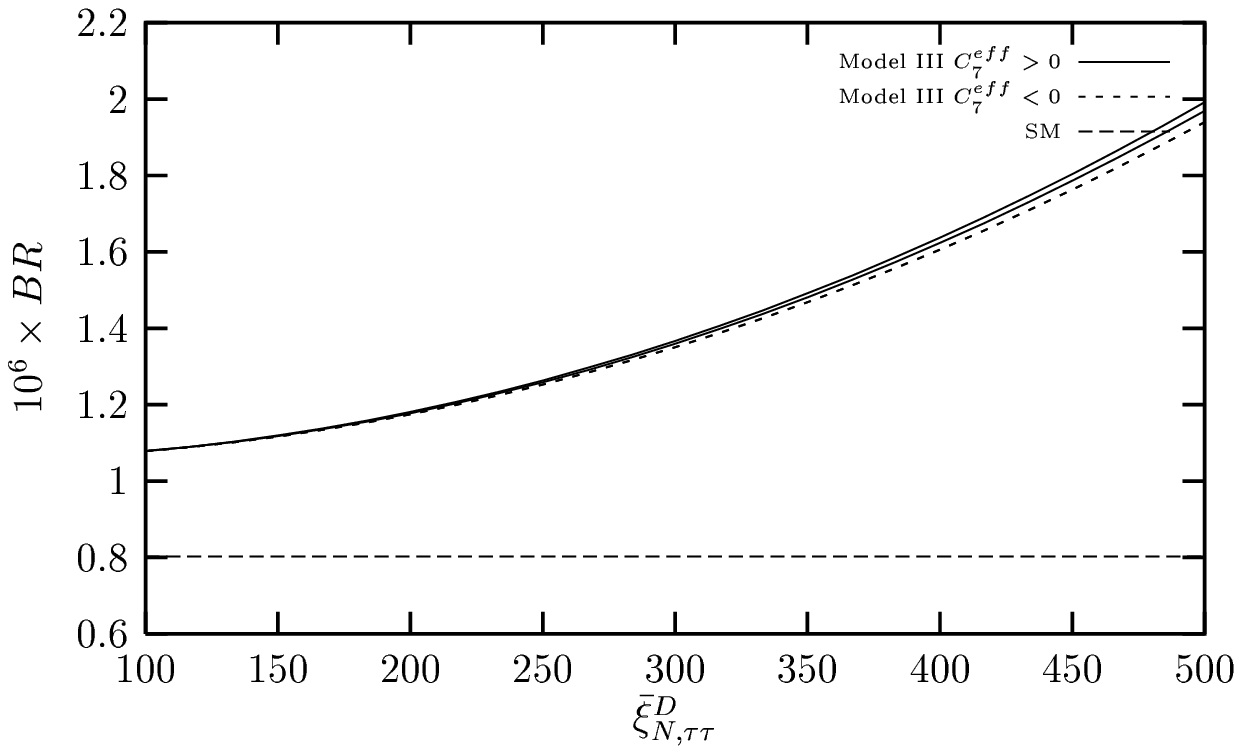}
\vskip -3.0truein
\caption[]{$BR$ as a function of $\bar{\xi}_{N,\tau\tau}^{D}$, for 
$\bar{\xi}_{N,bb}^{D}=40\, m_b$ in case of the ratio $|r_{tb}|<1$. }
\label{brtauRk1}
\end{figure}

\begin{figure}[htb]
\vskip -3.0truein
\centering
\epsfxsize=6.8in
\leavevmode\epsffile{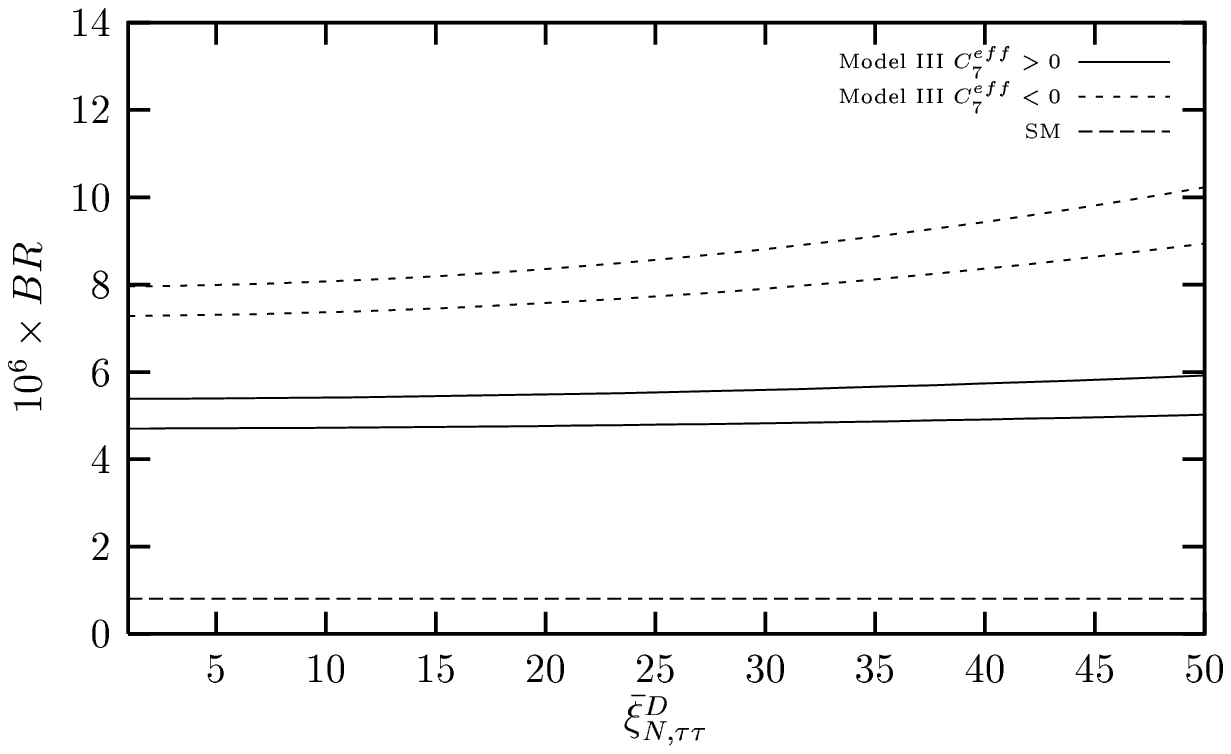}
\vskip -3.0truein
\caption[]{$BR$ as a function of $\bar{\xi}_{N,\tau\tau}^{D}$, for 
$\bar{\xi}_{N,bb}^{D}=3\, m_b$ in case of the ratio $r_{tb}>1$. }
\label{brtauRb1}
\end{figure}

Figures (\ref{brtauRk1}-\ref{brtauRb1}) represent the dependencies of 
the $BR$ on the parameter $\bar{\xi}_{N,\tau\tau}^{D}$ for $|r_{tb}|<1$ and 
$r_{tb}>1$ respectively. In $r_{tb}<1$ case, the $BR$ increases almost 
$1.5$ times compared to the one in the SM for large values of   
$\bar{\xi}_{N,\tau\tau}^{D}$, $\bar{\xi}_{N,\tau\tau}^{D}=500\, GeV$ 
(Fig. \ref{brtauRk1}). However, for $r_{tb}>1$, this enhancement is quite
high as shown in Fig. \ref{brtauRb1}. Even for small values of 
$\bar{\xi}_{N,bb}^{D}$ and $\bar{\xi}_{N,\tau\tau}^{D}$ there is a possible 
increase nearly (more than) one order of magnitude compared to the SM case 
for $C^{eff}_{7}>0$ ($C^{eff}_{7}<0$).     

Now we would like to summarize our results:
\begin{itemize}
\item There is a possible enhancement in the $BR$ at the order of magnitude
$\% \, 150$ for $|r_{tb}|<1$ in the model III compared to the one in the SM 
for large values of the model III parameters, $\bar{\xi}_{N,bb}^{D}=80\, m_b$
and  $\bar{\xi}_{N,\tau\tau}^{D}=500\, GeV$. The $BR$ is not so much 
sensitive to the model parameters given above. Further, the NHB effects 
become sizable with increasing values of $\bar{\xi}_{N,\tau\tau}^{D}$.
\item For $r_{tb}>1$, there is a considerable  enhancement at the one order
of magnitude  compared to the SM, even for the small values of 
$\bar{\xi}_{N,bb}^{D}$ and $\bar{\xi}_{N,\tau\tau}^{D}$. In this case, 
the $BR$ is larger and more sensitive the model parameters for 
$C^{eff}_{7}<0$ than the ones for $C^{eff}_{7}>0$. Note that the enhancement 
for the increasing values of
the $\bar{\xi}_{N,\tau\tau}^{D}$ is due to the dependence of the $BR$ on the NHB
effects.
\end{itemize}

{\Large{Acknowledgement}}

We would like to thank Liao Wei for his comments about the previous version
of this manuscript.
\newpage
{\bf \LARGE {Appendix}} \\
\begin{appendix}

\section{The operator basis }
The operator basis in the  2HDM (model III ) for our process  
is \cite{Chao,Grinstein2,misiak}
\begin{eqnarray}
 O_1 &=& (\bar{s}_{L \alpha} \gamma_\mu c_{L \beta})
               (\bar{c}_{L \beta} \gamma^\mu b_{L \alpha}), \nonumber   \\
 O_2 &=& (\bar{s}_{L \alpha} \gamma_\mu c_{L \alpha})
               (\bar{c}_{L \beta} \gamma^\mu b_{L \beta}),  \nonumber   \\
 O_3 &=& (\bar{s}_{L \alpha} \gamma_\mu b_{L \alpha})
               \sum_{q=u,d,s,c,b}
               (\bar{q}_{L \beta} \gamma^\mu q_{L \beta}),  \nonumber   \\
 O_4 &=& (\bar{s}_{L \alpha} \gamma_\mu b_{L \beta})
                \sum_{q=u,d,s,c,b}
               (\bar{q}_{L \beta} \gamma^\mu q_{L \alpha}),   \nonumber  \\
 O_5 &=& (\bar{s}_{L \alpha} \gamma_\mu b_{L \alpha})
               \sum_{q=u,d,s,c,b}
               (\bar{q}_{R \beta} \gamma^\mu q_{R \beta}),   \nonumber  \\
 O_6 &=& (\bar{s}_{L \alpha} \gamma_\mu b_{L \beta})
                \sum_{q=u,d,s,c,b}
               (\bar{q}_{R \beta} \gamma^\mu q_{R \alpha}),  \nonumber   \\  
 O_7 &=& \frac{e}{16 \pi^2}
          \bar{s}_{\alpha} \sigma_{\mu \nu} (m_b R + m_s L) b_{\alpha}
                {\cal{F}}^{\mu \nu},                             \nonumber  \\
 O_8 &=& \frac{g}{16 \pi^2}
    \bar{s}_{\alpha} T_{\alpha \beta}^a \sigma_{\mu \nu} (m_b R +
m_s L)  
          b_{\beta} {\cal{G}}^{a \mu \nu} \nonumber \,\, , \\  
 O_9 &=& \frac{e}{16 \pi^2}
          (\bar{s}_{L \alpha} \gamma_\mu b_{L \alpha})
              (\bar{\tau} \gamma^\mu \tau)  \,\, ,    \nonumber    \\
 O_{10} &=& \frac{e}{16 \pi^2}
          (\bar{s}_{L \alpha} \gamma_\mu b_{L \alpha})
              (\bar{\tau} \gamma^\mu \gamma_{5} \tau)  \,\, ,    \nonumber  \\
Q_1&=&   \frac{e^2}{16 \pi^2}(\bar{s}^{\alpha}_{L}\,b^{\alpha}_{R})\,(\bar{\tau}\tau ) 
\nnb  \\ 
Q_2&=&    \frac{e^2}{16 \pi^2}(\bar{s}^{\alpha}_{L}\,b^{\alpha}_{R})\,
(\bar{\tau} \gamma_5 \tau ) \nnb \\
Q_3&=&    \frac{g^2}{16 \pi^2}(\bar{s}^{\alpha}_{L}\,b^{\alpha}_{R})\,
\sum_{q=u,d,s,c,b }(\bar{q}^{\beta}_{L} \, q^{\beta}_{R} ) \nnb \\
Q_4&=&  \frac{g^2}{16 \pi^2}(\bar{s}^{\alpha}_{L}\,b^{\alpha}_{R})\,
\sum_{q=u,d,s,c,b } (\bar{q}^{\beta}_{R} \, q^{\beta}_{L} ) \nnb \\
Q_5&=&   \frac{g^2}{16 \pi^2}(\bar{s}^{\alpha}_{L}\,b^{\beta}_{R})\,
\sum_{q=u,d,s,c,b } (\bar{q}^{\beta}_{L} \, q^{\alpha}_{R} ) \nnb \\
Q_6&=&   \frac{g^2}{16 \pi^2}(\bar{s}^{\alpha}_{L}\,b^{\beta}_{R})\,
\sum_{q=u,d,s,c,b } (\bar{q}^{\beta}_{R} \, q^{\alpha}_{L} ) \nnb \\
Q_7&=&   \frac{g^2}{16 \pi^2}(\bar{s}^{\alpha}_{L}\,\sigma^{\mu \nu} \, 
b^{\alpha}_{R})\,
\sum_{q=u,d,s,c,b } (\bar{q}^{\beta}_{L} \, \sigma_{\mu \nu } 
q^{\beta}_{R} ) \nnb \\
Q_8&=&    \frac{g^2}{16 \pi^2}(\bar{s}^{\alpha}_{L}\,\sigma^{\mu \nu} 
\, b^{\alpha}_{R})\,
\sum_{q=u,d,s,c,b } (\bar{q}^{\beta}_{R} \, \sigma_{\mu \nu } 
q^{\beta}_{L} ) \nnb \\ 
Q_9&=&   \frac{g^2}{16 \pi^2}(\bar{s}^{\alpha}_{L}\,\sigma^{\mu \nu} 
\, b^{\beta}_{R})\,
\sum_{q=u,d,s,c,b }(\bar{q}^{\beta}_{L} \, \sigma_{\mu \nu } 
q^{\alpha}_{R} ) \nnb \\
Q_{10}&= & \frac{g^2}{16 \pi^2}(\bar{s}^{\alpha}_{L}\,\sigma^{\mu \nu} \, 
b^{\beta}_{R})\,
\sum_{q=u,d,s,c,b }(\bar{q}^{\beta}_{R} \, \sigma_{\mu \nu } q^{\alpha}_{L} )
\label{op1}
\end{eqnarray}
where $\alpha$ and $\beta$ are $SU(3)$ colour indices and 
${\cal{F}}^{\mu \nu}$ and ${\cal{G}}^{\mu \nu}$ are the field strength 
tensors of the electromagnetic and strong interactions, respectively. Note 
that there are also flipped chirality partners of these operators, which 
can be obtained by interchanging $L$ and $R$ in the basis given above in 
the model III. However, we do not present them here since corresponding  Wilson 
coefficients are negligible.
\section{The Initial values of the Wilson coefficients.}
For the sake of completeness we also give the initial values of the Wilson 
coefficients for the relevant process.  In the SM they are \cite{Grinstein2}
\begin{eqnarray}
C^{SM}_{1,3,\dots 6,11,12}(m_W)&=&0 \nonumber \, \, , \\
C^{SM}_2(m_W)&=&1 \nonumber \, \, , \\
C_7^{SM}(m_W)&=&\frac{3 x_t^3-2 x_t^2}{4(x_t-1)^4} \ln x_t+
\frac{-8 x_t^3-5 x_t^2+7 x_t}{24 (x_t-1)^3} \nonumber \, \, , \\
C_8^{SM}(m_W)&=&-\frac{3 x_t^2}{4(x_t-1)^4} \ln x_t+
\frac{-x_t^3+5 x_t^2+2 x_t}{8 (x_t-1)^3}\nonumber \, \, , \\ 
C_9^{SM}(m_W)&=&-\frac{1}{sin^2\theta_{W}} B(x_t) +
\frac{1-4 \sin^2 \theta_W}{\sin^2 \theta_W} C(x_t)-D(x_t)+
\frac{4}{9}, \nonumber \, \, , \\
C_{10}^{SM}(m_W)&=&\frac{1}{\sin^2\theta_W}
(B(x_t)-C(x_t))\nonumber \,\, , \\
C_{Q_i}^{SM}(m_W) & = & 0~~~ i=1,..,10~.
\end{eqnarray}
The initial values for the additional part due to charged Higgs bosons are 
\begin{eqnarray}
C^{H}_{1,\dots 6 }(m_W)&=&0 \nonumber \, , \\
C_7^{H}(m_W)&=& Y^2 \, F_{1}(y_t)\, + \, X Y \,  F_{2}(y_t) 
\nonumber  \, \, , \\
C_8^{H}(m_W)&=& Y^2 \,  G_{1}(y_t) \, + \, X Y \, G_{2}(y_t) 
\nonumber\, \, , \\
C_9^{H}(m_W)&=&  Y^2 \,  H_{1}(y_t) \nonumber  \, \, , \\
C_{10}^{H}(m_W)&=& Y^2 \,  L_{1}(y_t)  
\label{CH} \, \, , 
\end{eqnarray}
where 
\bea
X & = & \frac{1}{m_{b}}~~~\left(\bar{\xi}^{D}_{N,bb}+\bar{\xi}^{D}_{N,sb}
\frac{V_{ts}}{V_{tb}} \right) ~~,~~ \nnb \\
Y & = & \frac{1}{m_{t}}~~~\left(\bar{\xi}^{U}_{N,tt}+\bar{\xi}^{U}_{N,tc}
\frac{V^{*}_{cs}}{V^{*}_{ts}} \right) ~~,~~
\eea
and due to the neutral  Higgs bosons are \cite{gurer2}
\bea
C^{A^{0}}_{Q_{2}}((\bar{\xi}^{U}_{N,tt})^{3})  =  
\frac{\bar{\xi}^{D}_{N,\tau \tau}(\bar{\xi}^{U}_{N,tt})^{3}
m_{b} y_t}{32 \pi^{2}m_{A^{0}}^{2}m_{t} \Theta_1 (z_{A}) (y_t-1)^2} 
((y_t-1) (-\Theta_1 (z_{A})+
(y_t-1) z_A)+\Theta_1 (z_{A}) \ln y_t), \nnb
\eea
\bea
C^{A^{0}}_{Q_{2}}((\bar{\xi}^{U}_{N,tt})^{2}) & = & \frac{\bar{\xi}^{D}_{N,\tau\tau}(\bar{\xi}^{U}_{N,tt})^{2}
\bar{\xi}^{D}_{N,bb}}{32 \pi^{2}  m_{A^{0}}^{2}} 
\left(\frac{(1-2 y_t) \ln y_t}{y_t-1}+2 (1+\ln \Big{[} \frac{\Theta_1
(z_{A})}{z_{A}}\Big{]} )-\frac{y_t (x y +z_A)}{\Theta_1 (z_{A})} \right), \nnb
\eea
\bea
C^{A^{0}}_{Q_{2}}(\bar{\xi}^{U}_{N,tt}) &=& 
\frac{g^2\bar{\xi}^{D}_{N,\tau\tau}\bar{\xi}^{U}_{N,tt} m_b
x_t}{128 \pi^2 m_t } \Bigg{(} 2 z_A\frac{- x (\frac{y}{\Theta_2 (z_{A})}+
\frac{y_t}{\Theta_3(z_{A})} )
+\frac{y_t (x-1)}{-\Theta_3 (z_{A})+(x-y)(x_t-y_t)z_{A}}}{m^2_W} \nnb \\
&+& \Bigg{(}\frac{1}{m^2_{A^{0}}}
(-\frac{4 z_{A}}{\Theta_2 (z_{A})}+
\frac{2 (x(x_t+y_t)-2 y_t) z_{A}}{-\Theta_3(z_{A})}-\frac{2 (x_t(x-1)+(x+1) y_t)
z_{A}}{\Theta_3 (z_{A})+(x-y)(x_t-y_t)z_{A}} \nnb \\
 &+&\frac{ (y_t-x_t) z_{A}+x_t y_t (2 z_{A}-1)}{(x_t-1)(y_t-1) z_{A}} 
 -\frac{(4 x_t^3-7 y_t-4 x_t^2 (2+y_t)+x_t(5+8 y_t+\frac{y_t}{z_A}))
\ln x_t}{(x_t-1)^2 (x_t-y_t) } \nnb \\
&+& \frac{(x_t (\frac{y_t}{z_A}-1)-y_t)\ln
y_t}{(x_t-y_t)(y_t-1)^2} \nnb +  4 \ln \Big{[} \frac{\Theta_2
(z_A)}{z_A}\Big{]}\Bigg{)}\Bigg{)}, \nnb 
\eea
\bea
\lefteqn{C^{A^{0}}_{Q_{2}}(\bar{\xi}^{D}_{N,bb}) =
-\frac{g^2\bar{\xi}^{D}_{N,\tau\tau}\bar{\xi}^{D}_{N,bb}
}{64 \pi^2 m^2_{A^{0}} } \Bigg{(}
\frac{2 \Theta_3 (z_{A})-x_t ((x-2) y y_t-x_t (y_t-z_A))}
{\Theta_3(z_{A})}+2 \ln \Big{[}\frac{\Theta_3 (z_{A})}{z_A} 
\Big{]}}\nnb 
\\
& &- \frac{(y_t-2 x_t (y_t+1)+x^2_t(\frac{y_t}{z_A}+1) \ln x_t}{(x_t-1)
(x_t-y_t)}+ \frac{(x_t (1-2 y_t)+2 y_t (y_t-1)+x^2_t(\frac{y_t}{z_A}-1)) 
\ln y_t}{(y_t-1)(x_t-y_t)} \Bigg{)}, \nnb
\eea
\bea
\lefteqn{C^{H^{0}}_{Q_{1}}((\bar{\xi}^{U}_{N,tt})^{2}) = 
-\frac{g^2 (\bar{\xi}^{U}_{N,tt})^2 m_b m_{\tau}y_t
}{256 \pi^2 m^2_{H^{0}} m^2_W x_t } \Bigg{(}
-\frac{ \frac{4 x z_H}{\Theta_4 (z_{H})}+\frac{-1+4 y_t+y^2_t (2 \ln
y_t-3)}{(y_t-1)^3}}{\cos^2 \theta_W}+} \nnb
\\  & &       
2 \Big{(} \frac{2 z_H (-\Theta_4 (z_{H})(1-2 x) x_t+2 \Theta_1
(z_{H})x)}{\Theta_1 (z_{H})\Theta_4(z_{H})}-
\frac{(y_t-1)(1+x_t+y_t(x_t-3))+2 y_t (y_t-x_t)\ln y_t}{(y_t-1)^3}
 \Big{)} \Big{)} , \nnb 
\eea
\bea
\lefteqn{\!\!\!\!\!\!\!\!\!\!\!\!\!C^{H^{0}}_{Q_{1}}(\bar{\xi}^{U}_{N,tt}) 
=\frac{g^2 \bar{\xi}^{U}_{N,tt} \bar{\xi}^{D}_{N,bb} 
m_{\tau}}{64 \pi^2 m^2_{H^{0}} m_t } \Bigg{(}
\frac{y_t (2-x_t)-x_t}{y_t-1}
+2 x_t \ln \Big{[}\frac{\Theta_1 (z_{H})}{z_H} \Big{]}
 -\frac{(x_t (1-5 y_t)+2 y^2_t(1+x_t))\ln y_t}{(y_t-1)^2}}\nnb
\\  & &\!\!\!\! -\frac{z_H}
{\Theta_1 (z_{H})\Theta_4(z_{H})}
\Big{(} -x^2 x_t \Theta_5 (y_t-1)+x (x_t (-\Theta_4 (z_{H})+\Theta_5
(1+x-y)) (y_t-1) \nnb \\ & + & 2 y_t ( -\Theta_5 +2 y y_t))  +
(-x_t \Theta_6 (y_t-1)-2(1+y(y_t-1)) y_t)z_H \Big{)} \nnb \\ & + &
\frac{y_t ((y_t-1) (-\Theta_4 (z_{H})+z_H (1-y_t))+\Theta_4 (z_{H}) y_t 
\ln y_t )}{\cos^2 \theta_W \Theta_4 (z_{H})(y_t-1)^2}
 \Big{)}, \nnb
\eea
\bea
\lefteqn{ C^{H^0}_{Q_{1}}(g^4) =-\frac{g^4 m_b m_{\tau}}{512 \pi^2 m^2_{H^{0}} m^2_W } 
\Bigg{(}\frac{\frac{4 (x-1)}{\Theta_7 }+\frac{x_t(x_t (4-x_t)-3+2 x_t
(x_t-2) \ln x_t)}{(x_t-1)^3}}{\cos^2 \theta_W}}\nnb
\\  & &- \frac{4 (x (4+2 x_t-\frac{x_t y}{z_H})-2)}{\Theta_8}
-\frac{4 (x_t ( \frac{ y (2-x)}{z_H}+3+\Theta_7 )-4 x)+
2 \Theta_7 x_t \ln \Theta_7 )}{x_t \Theta_7 }\nnb \\ 
& & +\frac{2 (2-12 x_t+21 xt^2-12 xt^3+x_t^4+(2-4 x_t-2 x_t^2+6 x_t^3)
\ln x_t)}{(x_t-1)^3}-4 x_t (1+2 \ln \Big{[} \Theta_8 \Big{]}) \Big{)},\nnb
\eea
\bea
C^{h_0}_{Q_1}((\bar{\xi}^U_{N,tt})^3) &=&
-\frac{\bar{\xi}^D_{N,\tau\tau} (\bar{\xi}^U_{N,tt})^3 m_b y_t}
{64 \Theta_1 (z_h) m_{h^0}^2 m_t\pi^2 (-1+y_t)^3}
 ((-1+y_t) (\Theta_1 (z_h)(y_t+1)\nonumber 
\\
&+& 2 (2x-1) (y_t-1)^2 z_h)-
2 \Theta_1 (z_h) y_t\ln y_t ) ,\nonumber \\ 
C^{h_0}_{Q_1}((\bar{\xi}^U_{N,tt})^2)&=&
-\frac{1}{32m_{h^0}^2\pi^2} \bar{\xi}^D_{N,\tau\tau}  \bar{\xi}^D_{N,bb} 
(\bar{\xi}^U_{N,tt})^2 \Big{(}- \frac{\Theta_1 (z_h)+ y_t (x y-z_h)}
{\Theta_1 (z_h)} + \frac{(1-y_t+(-1+2y_t)\ln y_t}{-1+y_t} \nonumber
\\ 
&- & 2 \ln \Big{[}\frac{\Theta_1 (z_h)}{z_h}\Big{]} \Big{)},\nnb 
\eea
\bea
\lefteqn{ C^{h^0}_{Q_{1}}(\bar{\xi}^{U}_{N,tt}) =-\frac{g^2
\bar{\xi}^{D}_{N,\tau\tau}\bar{\xi}^{U}_{N,tt}x_t}{128 \pi^2 m^2_{h^{0}} m_t }         
\Bigg{(}\frac{x_t (8-9 y_t)-x_t^3 (y_t-2)+y_t (5 y_t-4)+x_t^2 (-4+2
y_t+y_t^2)}{\Theta_5 (x_t-1)}} \nnb 
\\ & & -\frac{y_t x_t (2-x_t-y_t)}{ z_h \Theta_5}
-\frac{4 z_h (-1+x (2+x_t))-2 x y x_t}{\Theta_2 (z_h)}+\frac{2 z_h (-2 y_t+x
(x_t+y_t))}{\Theta_3 (z_h)} \nnb 
\\ & &+\frac{2 z_h (x_t (x-1)+y_t (x+1))}
{-\Theta_3 (z_h)+(x-y) (x_t-y_t)z_h}+
\frac{4 (-1+x) x x_t y_{t}^{2} z_{h}}{(\Theta_4 (z_h) x_t - 
x (x_t-y_t) z_h)(\Theta_4 (z_h) x_t - y (x_t-y_t) z_h)}\nnb 
\\ & & 
+\Big{(} 4+\frac{(y_t-1)^2 (x_t^3 (3-10 y_t)+7 y_t^2-7 x_t
y_t(2+y_t)+3 x_t^2 (1+4 y_t+2 y_t^2))}{  (x_t-1)}\nnb \
\\ & & +\frac{y_t x_t  (-1+y_t)^2 (-\Theta_6 + 4 (x_t-y_t) y_t) 
}{z_h}\Big{)} \frac{\ln x_t}{\Theta_5^2}
\nnb \\ & & -
\frac{(x_t-1)^2 (-10 x_t y_t (y_t-1)+x_t^2 (2 y_t-1)+y_t^2 (4 y_t-5)-
\frac{x_t y_t}{z_h} \Theta_6 ) \ln y_t}{\Theta^2_5}-4 \ln \Big{[} 
\frac{\Theta_2 (z_h)}{z_h}\Big{]}\Big{)}, \nnb
\eea
\bea
\lefteqn{ C^{h^0}_{Q_{1}}(\bar{\xi}^{D}_{N,bb}) =-\frac{g^2
\bar{\xi}^{D}_{N,\tau\tau}\bar{\xi}^{D}_{N,bb}z_h}{64 \pi^2 m^2_W x_t }
\Bigg{(}\frac{(x_t^2 (\frac{y_t}{z_h}+1)+y_t-2 x_t (y_t+1)) \ln x_t}
{(x_t-1) (x_t-y_t)}} \nnb 
\\ & & +
\frac{(x_t^2 (\frac{y_t}{z_h}-1)+2 y_t (y_t-1)+x_t (1-2 y_t)) \ln y_t}
{(y_t-1) (x_t-y_t)}- 2 \ln \Big{[} \frac{\Theta_3 (z_h)}{z_h}\Big{]} \nnb 
\\  & & +x_t\frac{y_t (x_t+2 y(x-1))-z_h (x_t-2 y_t (y-1)+2 y
\frac{y_t}{x_t})+x (y y_t+2 z_h (y_t-1))}{\Theta_3 (z_h)} \Big{)},  
\label{NHB}
\eea
where
\bea
\Theta_1 (\omega) & = & ((1-y+y y_t) \omega -x (y y_t+\omega(1-y_t)) \nnb \\
\Theta_2 (\omega) & = & \Theta_1 (\omega, y_t\rightarrow x_t) \nnb \\
\Theta_3 (\omega) & = & (x_t (1-y)+y ) y_t \omega -
x x_t (y y_t+\omega(-1+y_t)) \nnb \\
\Theta_4 (\omega) & = & (y(1-y_t)+ y_t) \omega -
x (y y_t+\omega(-1+y_t)) \nnb \\
\Theta_5  & = & (-1+x_t)(x_t-y_t)(-1+y_t) \nnb \\
\Theta_6  & = & (-1+y) (y (-1+y_t)y_t)\nnb \\
\Theta_7  & = & \frac{(x_t+y (1-x_t)) z_h+x (z_h-x_t (y+z_h))}{x_t z_h} \nnb
\\
\Theta_8  & = & \frac{(1-y (1-x_t)) z_h-
x (x_t (y-z_h)+z_h)}{x_t z_h}
\eea
and
\begin{eqnarray}
& & x_t=\frac{m_t^2}{m_W^2}~~~,~~~y_t=
\frac{m_t^2}{m_{H^{\pm}}}~~~,~~~z_H=\frac{m_t^2}{m^2_{H^0}}~~~,~~~
z_h=\frac{m_t^2}{m^2_{h^0}}~~~,~~~ z_A=\frac{m_t^2}{m^2_{A^0}}~~~,~~~ 
\label{CoeffH}
\end{eqnarray}
The explicit forms of the functions $F_{1(2)}(y_t)$, $G_{1(2)}(y_t)$, 
$H_{1}(y_t)$ and $L_{1}(y_t)$ in eq.(\ref{CH}) are given as
\begin{eqnarray}
F_{1}(y_t)&=& \frac{y_t(7-5 y_t-8 y_t^2)}{72 (y_t-1)^3}+
\frac{y_t^2 (3 y_t-2)}{12(y_t-1)^4} \,\ln y_t \nonumber  \,\, , 
\\ 
F_{2}(y_t)&=& \frac{y_t(5 y_t-3)}{12 (y_t-1)^2}+
\frac{y_t(-3 y_t+2)}{6(y_t-1)^3}\, \ln y_t 
\nonumber  \,\, ,
\\ 
G_{1}(y_t)&=& \frac{y_t(-y_t^2+5 y_t+2)}{24 (y_t-1)^3}+
\frac{-y_t^2} {4(y_t-1)^4} \, \ln y_t
\nonumber  \,\, ,
\\ 
G_{2}(y_t)&=& \frac{y_t(y_t-3)}{4 (y_t-1)^2}+\frac{y_t} {2(y_t-1)^3} \, 
\ln y_t  \nonumber\,\, ,
\\
H_{1}(y_t)&=& \frac{1-4 sin^2\theta_W}{sin^2\theta_W}\,\, \frac{xy_t}{8}\,
\left[ \frac{1}{y_t-1}-\frac{1}{(y_t-1)^2} \ln y_t \right]\nonumber \\
&-&
y_t \left[\frac{47 y_t^2-79 y_t+38}{108 (y_t-1)^3}-
\frac{3 y_t^3-6 y_t+4}{18(y_t-1)^4} \ln y_t \right] 
\nonumber  \,\, , 
\\ 
L_{1}(y_t)&=& \frac{1}{sin^2\theta_W} \,\,\frac{x y_t}{8}\, 
\left[-\frac{1}{y_t-1}+ \frac{1}{(y_t-1)^2} \ln y_t \right]
\nonumber  \,\, .
\\ 
\label{F1G1}
\end{eqnarray}
Finally, the initial values of the coefficients in the model III are
\begin {eqnarray}   
C_i^{2HDM}(m_{W})&=&C_i^{SM}(m_{W})+C_i^{H}(m_{W}) , \nnb \\
C_{Q_{1}}^{2HDM}(m_{W})&=& \int^{1}_{0}dx \int^{1-x}_{0} dy \,
(C^{H^{0}}_{Q_{1}}((\bar{\xi}^{U}_{N,tt})^{2})+
 C^{H^{0}}_{Q_{1}}(\bar{\xi}^{U}_{N,tt})+
 C^{H^{0}}_{Q_{1}}(g^{4})+C^{h^{0}}_{Q_{1}}((\bar{\xi}^{U}_{N,tt})^{3}) \nnb
\\ & + &
 C^{h^{0}}_{Q_{1}}((\bar{\xi}^{U}_{N,tt})^{2})+
 C^{h^{0}}_{Q_{1}}(\bar{\xi}^{U}_{N,tt})+
 C^{h^{0}}_{Q_{1}}(\bar{\xi}^{D}_{N,bb})) , \nnb  \\
 C_{Q_{2}}^{2HDM}(m_{W})&=& \int^{1}_{0}dx \int^{1-x}_{0} dy\,
(C^{A^{0}}_{Q_{2}}((\bar{\xi}^{U}_{N,tt})^{3})+
C^{A^{0}}_{Q_{2}}((\bar{\xi}^{U}_{N,tt})^{2})+
 C^{A^{0}}_{Q_{2}}(\bar{\xi}^{U}_{N,tt})+
 C^{A^{0}}_{Q_{2}}(\bar{\xi}^{D}_{N,bb}))\nnb \\
C_{Q_{3}}^{2HDM}(m_W) & = & \frac{m_b}{m_{\tau} \sin^2 \theta_W} 
 (C_{Q_{1}}^{2HDM}(m_W)+C_{Q_{2}}^{2HDM}(m_W)) \nnb \\
C_{Q_{4}}^{2HDM}(m_W) & = & \frac{m_b}{m_{\tau} \sin^2 \theta_W} 
 (C_{Q_{1}}^{2HDM}(m_W)-C_{Q_{2}}^{2HDM}(m_W)) \nnb \\
C_{Q_{i}}^{2HDM} (m_W) & = & 0\,\, , \,\, i=5,..., 10.
\label{CiW}
\end{eqnarray}
Here, we present $C_{Q_{1}}$ and $C_{Q_{2}}$ in terms of the Feynmann
parameters $x$ and $y$ since the integrated results are extremely large.
Using these initial values, we can calculate the coefficients 
$C_{i}^{2HDM}(\mu)$ and $C^{2HDM}_{Q_i}(\mu)$ 
at any lower scale in the effective theory 
with five quarks, namely $u,c,d,s,b$ similar to the SM case 
\cite{Chao,misiak, buras,alil2}. For completeness,
in the following we give the  explicit expressions for $C_{7}^{eff}(\mu)$
and $C_{9}^{eff}(\mu)$. 
\begin{eqnarray}
C_{7}^{eff}(\mu)&=&C_{7}^{2HDM}(\mu)+ Q_d \, 
(C_{5}^{2HDM}(\mu) + N_c \, C_{6}^{2HDM}(\mu))\nonumber \, \, , \\
&+& Q_u\, (\frac{m_c}{m_b}\, C_{12}^{2HDM}(\mu) + N_c \, 
\frac{m_c}{m_b}\,C_{11}^{2HDM}(\mu)) \, \, ,
\label{C7eff}
\end{eqnarray}
where the LO  QCD corrected Wilson coefficient 
$C_{7}^{LO, 2HDM}(\mu)$ is given by
\begin{eqnarray} 
C_{7}^{LO, 2HDM}(\mu)&=& \eta^{16/23} C_{7}^{2HDM}(m_{W})+(8/3) 
(\eta^{14/23}-\eta^{16/23}) C_{8}^{2HDM}(m_{W})\nonumber \,\, \\
&+& C_{2}^{2HDM}(m_{W}) \sum_{i=1}^{8} h_{i} \eta^{a_{i}} \,\, , 
\label{LOwils}
\end{eqnarray}
and $\eta =\alpha_{s}(m_{W})/\alpha_{s}(\mu)$, $h_{i}$ and $a_{i}$ are 
the numbers which appear during the evaluation \cite{buras}. 

The Wilson coefficient $C_9^{eff}(\mu)$ is :
\begin{eqnarray} 
C_9^{eff}(\mu)&=& C_9^{2HDM}(\mu) \nonumber 
\\ &+& h(z,  s) \left( 3 C_1(\mu) + C_2(\mu) + 3 C_3(\mu) + 
C_4(\mu) + 3 C_5(\mu) + C_6(\mu) \right) \nonumber \\
&- & \frac{1}{2} h(1, s) \left( 4 C_3(\mu) + 4 C_4(\mu) + 3
C_5(\mu) + C_6(\mu) \right) \\
&- &  \frac{1}{2} h(0,  s) \left( C_3(\mu) + 3 C_4(\mu) \right) +
\frac{2}{9} \left( 3 C_3(\mu) + C_4(\mu) + 3 C_5(\mu) + C_6(\mu)
\right) \nonumber \,\, .
\label{eqC9eff}
\end{eqnarray}
Here  the functions $h(u, s)$ are given by
\begin{eqnarray}
h(u, s) &=& -\frac{8}{9}\ln\frac{m_b}{\mu} - \frac{8}{9}\ln u +
\frac{8}{27} + \frac{4}{9} x \\
& & - \frac{2}{9} (2+x) |1-x|^{1/2} \left\{\begin{array}{ll}
\left( \ln\left| \frac{\sqrt{1-x} + 1}{\sqrt{1-x} - 1}\right| - 
i\pi \right), &\mbox{for } x \equiv \frac{4u^2}{ s} < 1 \nonumber \\
2 \arctan \frac{1}{\sqrt{x-1}}, & \mbox{for } x \equiv \frac
{4u^2}{ s} > 1,
\end{array}
\right. \\
h(0,s) &=& \frac{8}{27} -\frac{8}{9} \ln\frac{m_b}{\mu} - 
\frac{4}{9} \ln s + \frac{4}{9} i\pi \,\, , 
\label{hfunc}
\end{eqnarray}
with $u=\frac{m_c}{m_b}$.

Finally, the Wilson coefficient $C_{10}(\mu)$ is the same as $C_{10}(m_W)$ 
and $C_{Q_1}(\mu )$, $C_{Q_2}(\mu)$  
are given by \cite{Chao}
\beq
C_{Q_i}(\mu )=\eta^{-12/23}\,C_{Q_i}(m_W)~,~i=1,2~. 
\eeq
\end{appendix}
\newpage

\newpage
\end{document}